\documentclass{article}[12pt]
\begin{document}
\begin{center}
{\Huge Thermodynamics arising from Tsallis' thermostatistics\\}
\vspace{0.5cm}
{\Large F. Q. Potiguar\footnote{potiguar@fisica.ufc.br}, U. M. S. Costa\\}
\vspace{0.5cm}
{\em Universidade Federal do Cear\'a, Departamento de F\'\i sica, Campus do Pici, 60455-760, Fortaleza, Cear\'a, Brasil\\}
\vspace{1.0cm}
\section*{Abstract}
\end{center}
\noindent
We show, in two different ways, that the Tsallis' partition function and its derivatives are related to thermodynamic quantities such as entropy, internal energy, etc., in the same way as in Boltzmann-Gibbs' formalism, with the Lagrange multiplier $\beta^{BG}$ replaced by its value $\frac{1}{k_BT}$. They are obtained within the finite heat bath canonical ensemble approach. Furthermore, we discuss the meaning of the Lagrange multiplier of the generalized framework, $\beta^T$, and show that the entropy found here is just the R\`enyi entropy plus a definite constant.\\

PACs:05.20.-y, 05.70.-a


\newpage
\section{Introduction}
\noindent
One of the main achievements of the Boltzmann-Gibbs (BG) statistics is the microscopic derivation of thermodynamical quantities of a system, e.g., internal energy.\\
Specifically, in the canonical ensemble approach, where we study a system in thermal equilibrium with a much larger system, called heat bath, we have a quantity which plays a central role in this derivation, the system's partition function, $Z^{BG}_1$ \cite{Khinchin,Reif,Zemansky}.\\
A note must be said about the notation we use in this paper. In the following analysis we will have three systems: the studied, the heat bath and the composite (studied + bath). All quantities belonging to these three systems will be labeled by $1$, $2$, and $0$, respectively. When we deal with any function which is present in both frameworks, BG and Tsallis, we label it with a superscript, $BG$ or $T$, respectively.\\
The thermodynamical quantities are all given as functions of the natural logarithm of $Z^{BG}_1$ or as functions of its derivatives. The internal energy of the system is given by:
\begin{equation}
\label{internal-energy-01}
\left<E_1\right>=-\frac{\partial lnZ^{BG}_1}{\partial\beta^{BG}},
\end{equation}
where $\beta^{BG}=\frac{1}{k_BT}$, and $T$ is the equilibrium temperature set by the bath. Another quantity which has a macroscopic interpretation is the element of generalized work done {\em on} the system in a quasi-static process. It is given by:
\begin{equation}
\label{work-01}
\left<dW_1\right>=-\frac{1}{\beta^{BG}}\frac{\partial lnZ^{BG}_1}{\partial X_1}dX_1,
\end{equation}
where $X_1$ is the external parameter which characterizes the studied system. A third important quantity is the entropy of the system, which is given by:
\begin{equation}
\label{entropy-01}
S_1^{BG}=k_B(lnZ^{BG}_1+\beta^{BG}\left<E_1\right>).
\end{equation}
The Helmholtz free energy of this system follows from (\ref{entropy-01}) and is written as:
\begin{equation}
\label{Helmholtz-energy-01}
F_1=-\frac{1}{\beta^{BG}}lnZ^{BG}_1.
\end{equation}
In 1988, Tsallis \cite{Tsa88} proposed his non-extensive statistical mechanics. This formalism has been successfully applied to a variety of situations \cite{Tsa99}. For a regular updated list, see \cite{lista}. Some of the fields of applications are Levy flights \cite{ZaA95,PrT99}, turbulence \cite{Bec02}, and particle physics \cite{Bec02,RWW01}. Interpretations for the generalization parameter $q$ were also proposed, and all of them relate $q$ with the fluctuation of other important physical property in the context applied \cite{Ola01,WiW00}. \\
We follow here the path initiated by Plastino {\em et al.} \cite{PlasPlas94} and Almeida \cite{Alm01}, which we call finite heat bath canonical ensemble. It is a natural consequence of the structure function formalism \cite{Khinchin}.\\
This particular approach to Tsallis' statistics has been developed recently. In \cite{Adib02-01}, the generalized statistics was derived by a scale invariance argument, \cite{Andr02} deals with the connection with the Nos\'e-Hoover thermostat \cite{Nose,Hoover}, \cite{Arin02} made the connection of the present picture with the dynamical derivation of Beck \cite{Beck01}, \cite{Alm02} showed that Tsallis' entropy has an additive variant (and is connected with the entropy found here), in \cite{PoC02} the fluctuation of a system's energy in the canonical ensemble was calculated, and in \cite{Adib02-02} a numerical analysis of such finite systems was performed.\\
We propose the same problem as in the BG description: how are the ensemble averages of a system related to the system's partition function in Tsallis' framework? The Tsallis' partition function reads:
\begin{equation}
\label{partition-function}
Z^T_1=\int[1-(q_2-1)\beta^* E_1]^{\frac{1}{q_2-1}},
\end{equation}
where $q_2$ is associated to the number of degrees of freedom of the heat bath, $E_1$ is the system's bath and $\beta^*$ is a constant parameter. Although it looks like, this last parameter is not the Lagrange multiplier used in the optimization problem of Tsallis' entropy, $\beta^T$, as stated in \cite{TMP98}. For this reason we denote it with a star. In our discussion we will see where this difference arises in the calculations. We will obtain the microscopic formulae in two different ways.\\
Abe \cite{Abe01} has derived thermodynamics from Tsallis' statistics from purely macroscopic considerations which led him to modified thermodynamical laws obeyed by Tsallis' entropy form. We will see that these changes are not necessary, although we do not have the physical temperature as the proper integrating factor in the second law of thermodynamics for Tsallis' entropy. The R\'enyi entropy is the one with the physical temperature as the integrating factor in the second law, as found by Abe.\\
Also, we will see that no $q$-averages \cite{TMP98} or $q$-logarithms \cite{Cur92} are necessary for such a derivation to be made. This is another important feature of our work. We will use the distribution which arises from the optimization problem with the usual expectation value.\\
We organize this paper as follows: in section II, a description of the finite bath canonical ensemble is given. Section III contains our derivations. Section IV is reserved for a discussion of the various parameters introduced in the generalized context. Section V presents the discussion on the Tsallis' entropy and its role in our approach. Finally, in section VI we present our conclusions.\\

\section{Brief description of the finite heat bath picture}
\noindent
Let us consider a general system $i$ with hamiltonian $H_i(q,p)<E_i$. This system is characterized by a single external parameter $X_i$ (the extension of the following analysis to more parameters is immediate). The volume defined by the condition above, in the system's phase space, is defined as:
\begin{equation}
\label{phase-space-volume}
V_i(E_i)=\int_{H_i(q,p)<E_i}dV_i=C_i(X_i)E_i^{n_i},
\end{equation}
where $C_i$ is a coefficient which we assume to bear all the dependence on $X_i$ of this volume, and $n_i$ is a quantity proportional to the degrees of freedom of the system $i$. Its structure function, or its density of states, is defined as the area that encloses the above quantity and is defined as:
\begin{equation}
\label{structure-function}
\Omega_i(E_i)=\frac{d}{dE_i}V_i(E_i)=n_iC_i(X_i)E_i^{n_i-1}.
\end{equation}
The definition of the parameter $\beta_i$ is:
\begin{equation}
\label{beta-01}
\beta_i=\frac{d}{dE_i}ln\Omega_i(E_i).
\end{equation}
Almeida \cite{Alm01} has shown that for a system in contact with a heat bath to present a distribution function given as a power law, the following condition must be fulfilled:
\begin{equation}
\label{finite-bath-condition}
\frac{d}{dE_i}\left[\frac{1}{\beta_i(E_i)}\right]=q_i-1.
\end{equation}
It is worth to point out that this relation enables us to calculate the value of $q$ for any system, isolated or not, provided we know the dependence of $V_i(E_i)$ on $E_i$.\\
From equation (\ref{finite-bath-condition}) we have the general value of the parameter $q_i$:
\begin{equation}
\label{q-values}
q_i=\frac{n_i}{n_i-1}.
\end{equation}
From this relation we can rewrite equations (\ref{phase-space-volume}) and (\ref{structure-function}) as functions of $q_i$:
\begin{equation}
\label{phase-space-volume-q}
V_i(E_i)=C_i(X_i)E_i^{\frac{q_i}{q_i-1}},
\end{equation}
\begin{equation}
\label{structure-function-q}
\Omega_i(E_i)=\frac{q_i}{q_i-1}C_i(X_i)E_i^{\frac{1}{q_i-1}}.
\end{equation}
It is known \cite{Khinchin} that the energy distribution function of the system is given by:
\begin{equation}
\label{energy-distribution}
P_1(E_1)dE_1=\frac{\Omega_2(E_0-E_1)\Omega_1(E_1)}{\Omega_0(E_0)}dE_1.
\end{equation}
With the help of equation (\ref{structure-function-q}) we get:
\[
P_1(E_1)dE_1=\frac{\Omega_2(E_0)}{\Omega_0(E_0)}\left(1-\frac{E_1}{E_0}\right)^{\frac{1}{q_2-1}}\Omega_1(E_1)dE_1.
\]
If we identify:
\begin{equation}
\label{partition-function-01}
Z^T_1=\frac{\Omega_0(E_0)}{\Omega_2(E_0)},
\end{equation}
as the partition function and:
\begin{equation}
\label{beta-02}
\beta^*(q_2-1)=\frac{1}{E_0},
\end{equation}
we arrive at the Tsallis' power law canonical distribution:
\begin{equation}
\label{Tsallis-distribution}
P_1(E_1)=\frac{1}{Z^T_1}\left[1-(q_2-1)\beta^*E_1\right]^{\frac{1}{q_2-1}}dV_1.
\end{equation}
The average value of an energy dependent property $\lambda(E_1)$ is defined by:
\begin{equation}
\label{average-value}
\left<\lambda\right>=\frac{1}{Z^T_1}\int[1-(q_2-1)\beta^*E_1]^{\frac{1}{q_2-1}}\lambda dV_1,
\end{equation}
Equation (\ref{beta-02}) seems to be only a convenient choice of the parameter $\beta^*$, which appears from the optimization problem of Tsallis' entropy \cite{Tsa88}. But it is easy to show that this is the actual value of $\beta^*$ that gives the correct average energy $\left<E_1\right>$ as given by the energy equipartition theorem, which was shown to be valid in this context \cite{alm02-01}. In this reference, the authors suggested that the temperature is proportional to the energy per degree of freedom of a system:
\begin{equation}
\label{temperature-definition}
k_BT=\frac{E_0}{q_0/(q_0-1)}=\frac{\left<E_1\right>}{q_1/(q_1-1)}=\frac{\left<E_2\right>}{q_2/(q_2-1)}.
\end{equation}
This fact has already been pointed out in \cite{Reif,Prosper93}.\\
Let us calculate $\left<E_1\right>$, using (\ref{temperature-definition}):
\[
\frac{q_1}{q_1-1}\frac{q_0-1}{q_0}E_0=\frac{1}{Z^T_1}\int\left[1-(q_2-1)\beta^*E_1\right]^{\frac{1}{q_2-1}}E_1dV_1.
\]
This integral, and all others to come, must be evaluated bearing in mind the cut-off condition, $E_1\leq E_0$ \cite{PlasPlas94}, even if $q_2-1<0$, since this is a physical requirement for any case. This calculation usually yields beta functions. Hence, we arrive at:
\[
\frac{q_1}{q_1-1}\frac{q_0-1}{q_0}E_0=\left[(q_2-1)\beta^*\right]^{-1}\frac{B\left(\frac{q_1}{q_1-1}+1,\frac{q_2}{q_2-1}\right)}{B\left(\frac{q_1}{q_1-1},\frac{q_2}{q_2-1}\right)},
\]
which gives:
\[
\beta^*=\frac{1}{(q_2-1)E_0}.
\]
where $B(a,b)$ is the beta function of $a$ and $b$. This result was also found in \cite{Alm01}.\\
If we note that the parameters $q$ are related by the extensivity of the degrees of freedom \cite{alm02-01}:
\begin{equation}
\label{qs-relation}
\frac{q_0}{q_0-1}=\frac{q_1}{q_1-1}+\frac{q_2}{q_2-1}.
\end{equation}
The BG canonical ensemble can be recovered if we make the $q_2\rightarrow1$ limit. This is equivalent to make the number of degrees of freedom of the heat bath, and consequently of the whole system $\frac{q_0}{q_0-1}$, infinite. That is the reason why we call this formalism finite heat bath canonical ensemble.\\
What we propose here is to obtain the thermodynamic functions, such as internal energy and work, from the partition function $Z^T_1$:
\begin{equation}
\label{partition-function-02}
Z^T_1=\frac{q_0}{q_0-1}\frac{q_2-1}{q_2}C_1(X_1)E_0^{\frac{q_1}{q_1-1}}=\frac{q_0}{q_0-1}\frac{q_2-1}{q_2}V_1(E_0).
\end{equation}
We will show that this is a simple task, and that the relations obtained are the same as those obtained from the exponential distribution.

\section{Thermodynamics relations}
\noindent
We propose two different ways of deriving these relations. In both of them, we will see that the logarithm of $Z^T_1$ naturally appears in the calculations. Then, we do not need to replace it for $q$-averages \cite{TMP98} or $q$-logarithm \cite{Cur92} in order to obtain them.\\

\subsection{The first derivation}
\noindent
This path is motivated by the observation that the derivative of the integral form of the partition function does not have a simple form dependent on the partition function. This mathematical inconvenience can be directly viewed from the next equation. From (\ref{partition-function-01}), it can be shown that the system's partition function can be written as:
\[
Z^T_1=\int[1-(q_2-1)\beta^*E_1]^{\frac{1}{q_2-1}}\Omega_1(E_1)dE_1.
\]
Since the argument of $Z^T_1$ has the power $\frac{1}{q_2-1}$, any derivative of the first will give negative power of the argument, and no relation can be made with any observable quantity.\\
Therefore, we start with the definition of the following function:
\begin{equation}
\label{generator-function-01}
G_1=\int[1-(q_2-1)\beta^*E_1]^{\frac{q_2}{q_2-1}}dV_1=\int\left(1-\frac{q_0-1}{q_0}\frac{E_1}{k_BT}\right)^{\frac{q_2}{q_2-1}}dV_1.
\end{equation}
This is, indeed, not a casual choice of this function. We can relate it to the geometry of the phase space with the use of equations (\ref{phase-space-volume-q}) and (\ref{beta-02}). After a simple algebra, we can show that:
\begin{equation}
\label{generator-function-02}
G_1=\frac{V_0(E_0)}{V_2(E_0)}=V_1(E_0),
\end{equation}
which is the accessible volume of the studied system evaluated in $E_1=E_0$. From this result, we see that $G_1$ is proportional to the partition function $Z_1^T$:
\begin{equation}
\label{generator-function-03}
G_1=\frac{q_2}{q_2-1}\frac{q_0-1}{q_0}Z^T_1.
\end{equation}
It can also be directly calculated using equations (\ref{average-value}), (\ref{beta-02}), and (\ref{temperature-definition}).\\
Let us start our analysis with the calculation of the derivative of $G_1$ relating to the factor $\frac{1}{k_BT}$, using equation (\ref{generator-function-02}):
\[
\frac{\partial G_1}{\partial(1/k_BT)}=-\frac{q_2}{q_2-1}\frac{q_0-1}{q_0}\int\left(1-\frac{q_0-1}{q_0}\frac{E_1}{k_BT}\right)^{\frac{1}{q_2-1}}E_1
dV_1.
\]
Bearing in mind the average value definition (\ref{average-value}), we arrive at:
\begin{equation}
\label{internal-energy}
\left<E_1\right>=-\frac{\partial lnZ^T_1}{\partial(1/k_BT)}.
\end{equation}
This is the same relation between the internal energy of the system and its partition function, of course now we do not have $\beta^{BG}$ as the important variable, since we did not extremize the BG entropy, but we still have its actual value which is $\frac{1}{k_BT}$.\\
The second quantity to be determined is: $\frac{\partial G_1}{\partial X_1}$. Again, using equation (\ref{generator-function-02}), we have:
\[
\frac{\partial G_1}{\partial X_1}=-\frac{q_2}{q_2-1}\frac{q_0-1}{q_0}\frac{1}{k_BT}\int\left(1-\frac{q_0-1}{q_0}\frac{E_1}{k_BT}\right)^{\frac{1}{q_2-1}}\frac{\partial E_1}{\partial X_1}dV_1.
\]
Using (\ref{average-value}), we have:
\begin{equation}
\label{generalized-force}
\left<\frac{\partial E_1}{\partial X_1}\right>=-k_BT\frac{\partial lnZ^T_1}{\partial X_1}.
\end{equation}
This is the generalized force applied to the system. Hence, the work done on the system can be regarded as:
\begin{equation}
\label{generalized-work}
\left<dW\right>=\left<\frac{\partial E_1}{\partial X_1}\right>dX_1=-k_BT\frac{\partial lnZ^T_1}{\partial X_1}dX_1.
\end{equation}
Since we have the work done on the system in a quasi-static process, let us calculate the change in the internal energy during this process:
\[
d\left<E_1\right>-\left<dW_1\right>=d\left<E_1\right>+k_BT\frac{\partial lnZ^T_1}{\partial X_1}dX_1.
\]
Since $lnZ^T_1$ is a function of both $X_1$ and $\frac{1}{k_BT}$, we have, using (\ref{internal-energy}):
\[
\frac{1}{k_BT}(d\left<E_1\right>-\left<dW_1\right>)=
\]
\[dlnZ^T_1+\frac{1}{k_BT}d\left<E_1\right>+\left<E_1\right>d\left(\frac{1}{k_BT}\right),
\]
and we arrive at:
\[
d\left<E_1\right>-\left<dW_1\right>=Td\left[k_B\left(lnZ^T_1+\frac{\left<E_1\right>}{k_BT}\right)\right].
\]
From the first law of thermodynamics, we identify the right hand side of this equation as the infinitesimal quantity of heat rejected to the bath in this process, $\left<dQ_1\right>$. From the second law of thermodynamics, we related this heat with the entropy change in the system during this process: $\left<dQ_1\right>=TdS_1$. Then, the entropy of this system can be written as:
\begin{equation}
\label{entropy}
S_1=k_B\left(lnZ^T_1+\frac{\left<E_1\right>}{k_BT}\right).
\end{equation}
The definition of the Helmholtz free energy in thermodynamics is $F=\left<E\right>-TS$ for a general system, then we can readily obtain it from (\ref{entropy}) as:
\begin{equation}
\label{Helmholtz-energy-02}
F_1=-k_BTlnZ^T_1.
\end{equation}
We notice that all four relations, (\ref{internal-energy}), (\ref{generalized-work}), (\ref{entropy}), (\ref{Helmholtz-energy-02}), have the same form as in BG statistics.\\

\subsection{The second derivation}
\noindent
We will start this time from the direct definitions of internal energy and generalized work, and relate them with the partition function. This is the most common derivations in textbooks \cite{Khinchin,Reif}.\\
First, let us consider the internal energy, $\left<E_1\right>$, in the sense of (\ref{average-value}):
\[
\left<E_1\right>=\frac{1}{Z^T_1}\int\left(1-\frac{q_0-1}{q_0}\frac{E_1}{k_BT}\right)^{\frac{1}{q_2-1}}E_1dV_1.
\]
This is rewritten as:
\[
\left<E_1\right>=-\frac{q_0}{q_0-1}\frac{q_2-1}{q_2}\frac{1}{Z^T_1}\frac{\partial}{\partial(1/k_BT)}\int\left(1-\frac{q_0-1}{q_0}\frac{E_1}{k_BT}\right)^{\frac{q_2}{q_2-1}}dV_1.
\]
This integral is nothing less than the function $G_1$ defined in the previous section. Hence, using relation (\ref{generator-function-03}), we arrive at:
\begin{equation}
\left<E_1\right>=-\frac{\partial lnZ^T_1}{\partial(1/k_BT)},
\end{equation}
which is the same as before.\\
The generalized force is calculated as:
\[
\left<\frac{\partial E_1}{\partial X_1}\right>=\frac{1}{Z^T_1}\int\left(1-\frac{q_0-1}{q_0}\frac{E_1}{k_BT}\right)^{\frac{1}{q_2-1}}\frac{\partial E_1}{\partial X_1}dV_1.
\]
Rewriting this expression as:
\[
\left<\frac{\partial E_1}{\partial X_1}\right>=-\frac{q_0}{q_0-1}\frac{q_2-1}{q_2}\frac{k_BT}{Z^T_1}\frac{\partial}{\partial X_1}\int\left(1-\frac{q_0-1}{q_0}\frac{E_1}{k_BT}\right)^{\frac{q_2}{q_2-1}}dV_1.
\]
Again, using expression (\ref{generator-function-03}):
\begin{equation}
\left<\frac{\partial E_1}{\partial X_1}\right>=-k_BT\frac{\partial lnZ^T_1}{\partial X_1},
\end{equation}
which is the same as before.\\
Since the work and the internal energy expression are the same, we can directly write for the entropy, following the same path before, the next relation:
\begin{equation}
S_1=k_B\left(lnZ^T_1+\frac{\left<E_1\right>}{k_BT}\right).
\end{equation}
It also follows the Helmholtz free energy:
\begin{equation}
F_1=-k_BTlnZ^T_1.
\end{equation}
We have obtained an entropy, (\ref{entropy}), which is the sum of the natural logarithm of the partition function with the ratio $\frac{\left<E_1\right>}{T}$, which is equal $k_B\frac{q_1}{q_1-1}$. The first question which may be put is: which entropy is this? Since we are dealing with an approach to Tsallis' statistics, we should have obtained, at first sight, an expression like:
\begin{equation}
\label{Tsallis-entropy}
S_1^T=k_B\frac{1-\int P_1^{q{_2}}dV_1}{q_2-1}
\end{equation}
but it is clear that it is not possible to write Tsallis' entropy as a logarithm of the partition function. Then we rewrite $S_1$ as:
\[
S_1=k_BlnZ_1^T+k_B\frac{q_1}{q_1-1}.
\]
Therefore, we see that it is the form of the additive variant obtained by Almeida in \cite{Alm02}. Besides this remarkable fact, we can still investigate the role played by $S_1^T$ and $\beta^T$ in thermodynamics. Before answering this question, we will discuss the roles of the several $\beta$ parameters and the Tsallis' Lagrange multiplier $\beta^T$, as we stated earlier.\\

\section{The parameters $\beta_i$, $\beta^*$, and the Lagrange multiplier $\beta^T$}
\noindent
The thermodynamic relations for the work, internal energy and entropy as functions of the system's partition function are recovered and they do not depend explicitly on $q$. The parameter $\beta^{BG}$, which appears in the BG relations, is now replaced by its ordinary value $\frac{1}{k_BT}$, and we treat this factor as the important one. In this new context, $\beta^*$ has the following relation with the temperature:
\begin{equation}
\label{beta-T-relation}
\beta^*=\frac{1}{q_2-1}\frac{q_0-1}{q_0}\frac{1}{k_BT}
\end{equation}
which reduces to the original form when $q_2\rightarrow1$, since $q_0\rightarrow1$ as well, see equation (\ref{qs-relation}).\\
In other papers on the derivation of thermodynamics relations, $\beta^*$ is directly conncted to the physical temperature, $T$ \cite{Rama00,Abe00-02,Martinez00}. This is not correct under the finite heat bath picture. Hence, we use, now, the energy per degree of freedom as the physical temperature, and we can obtain all the previous relations.\\
Mathematically, $\beta^{BG}$ is introduced as a parameter contained in the approximation implied by the infinite bath condition \cite{Khinchin}, and it accounts for the bath's infinite energy. In information theory, it is the Lagrange multiplier associated with the mean internal energy constraint of the optimization problem of the entropy. Here, from equation (\ref{beta-02}), we see that $\beta^*$ and the parameter associated with the system, equation (\ref{finite-bath-condition}) with $i=1$, are different quantities. The former is a constant in equilibrium while the last ones are random quantities related to the temperature in an average way \cite{alm02-01}:
\begin{equation}
\label{beta-03}
k_BT=\frac{1}{q_i}\left<\frac{1}{\beta_i}\right>
\end{equation}
We did not mentioned the Tsallis' Lagrange multiplier $\beta^T$. In order to find its place, let us review the optimization problem \cite{Tsa88}. We must extremize Tsallis' entropy $S_1^T$, (\ref{Tsallis-entropy}), subjected to the two usual constraints, constant mean internal energy and normalization of the probability distribution $p_1$:
\begin{equation}
\label{mean-energy-constraint}
\left<E_1\right>=\int P_1E_1dV_1
\end{equation}
\begin{equation}
\label{normalization}
\int P_1dV_1=1.
\end{equation}
Introducing the Lagrange multipliers $\beta^T$ and $-\alpha$ for the former and the last constraints respectively, and applying the usual optimization technique, we arrive at the following equation for $P_1$:
\[
P_1^{q_2-1}=\frac{q_2-1}{q_2}\alpha-\frac{q_2-1}{q_2}\beta^TE_1.
\]
We rewrite this relation as follows:
\[
P_1^{q_2-1}=\frac{q_2-1}{q_2}\alpha\left[1-(q_2-1)\frac{\beta^T}{\alpha(q_2-1)}E_1\right].
\]
Now we define $\beta^*$ as:
\[
\beta^*=\frac{1}{(q_2-1)}\frac{\beta^T}{\alpha},
\]
and we arrive at:
\begin{equation}
P_1=\left[\frac{(q_2-1)\alpha}{q_2}\right]^{\frac{1}{q_2-1}}\left[1-(q_2-1)\beta^*E_1\right]^{\frac{1}{q_2-1}}.
\end{equation}
We see, as we argued before, that $\beta^*$ is not the multiplier $\beta^T$, it is a function of both multipliers. Because of this fact, we can foresee that it must be related to the partition function. This last function is defined by equation (\ref{normalization}):
\[
\left[\frac{(q_2-1)\alpha}{q_2}\right]^{\frac{1}{q_2-1}}=\frac{1}{Z^T_1}=\frac{1}{\int\left[1-(q_2-1)\beta^*E_1\right]^{\frac{1}{q_2-1}}dV_1}.
\]
Now we can write:
\begin{equation}
\label{alpha-definition}
(q_2-1)\alpha=\frac{q_2}{{Z^T_1}^{q_2-1}},
\end{equation}
and conclude that $\beta^*$ can be finally written as:
\begin{equation}
\label{beta-star}
\beta^*=\frac{{Z^T_1}^{q_2-1}}{q_2}\beta^T
\end{equation}
This relation is analogous to the one found in the optimization problem of $S_1^T$ with $q$ normalized average energy constraint \cite{TMP98}. The relation of $\beta^T$ with the temperature can be obtained using equation (\ref{beta-T-relation}). The former is given by:
\begin{equation}
\label{multiplier-T-relation}
\beta^Tk_BT=\frac{q_2}{q_2-1}\frac{q_0-1}{q_0}\frac{1}{{Z^T_1}^{q_2-1}},
\end{equation}
which is different from unity if we are not in the infinite bath limit, $q_2\rightarrow1$. Recalling the generating function relation to the partition function (\ref{generator-function-01}), we write the following equation:
\[
\beta^Tk_BT=\frac{G_1}{{Z^T_1}^{q_2}},
\]
which is nothing less than:
\[
\beta^Tk_BT=\int p_1^{q_2}dV_1.
\]
This integral is present in Tsallis' entropy formula. Hence, we rewrite the last as a function of the Lagrange multiplier:
\begin{equation}
\label{entropy-beta-relation}
\frac{S_1^T}{k_B}=\frac{1}{q_2-1}(1-\beta^Tk_BT)=\frac{1}{q_2-1}\left(1-\frac{q_2}{q_2-1}\frac{q_0-1}{q_0}\frac{1}{{Z^T_1}^{q_2-1}}\right).
\end{equation}
We see that Tsallis' entropy can be explicitly written in terms of the multiplier $\beta^T$.

\section{The problem of the entropy}
We now pass on to the matter of answering the question posed at the end of section III.B. Suppose we have a closed system in thermal equilibrium with another one at temperature $T$. This studied system is characterized by the external parameter $X_1$. Then we know that its entropy $S_1$ is a function of its internal energy $\left<E_1\right>$ and this parameter $X_1$. Since the entropy is an exact differential, we can write the next relation:
\begin{equation}
\label{entropy-differential}
dS_1=\frac{\partial S_1}{\partial\left<E_1\right>}d\left<E_1\right>+\frac{\partial S_1}{\partial X_1}dX_1.
\end{equation}
We now suppose that this entropy is given by Tsallis' one (\ref{entropy-beta-relation}). Since the above relation shows derivatives of the entropy, we calculate the derivative of (\ref{Tsallis-entropy}) with respect to a general quantity $Y$. After a simple algebra, we obtain:
\begin{equation}
\label{entropy-derivative-01}
\frac{\partial S_1^T}{\partial Y}=\beta^Tk_B^2T\frac{\partial lnZ^T_1}{\partial Y}.
\end{equation}
Now we have an interesting fact. The entropy itself is not a function of $lnZ^T_1$, but the derivatives of both are connected by a direct relation. We can explore this fact, together with the derivatives of $lnZ^T_1$.\\
First, we choose $Y=\left<E_1\right>$:
\[
\frac{\partial S_1^T}{\partial \left<E_1\right>}=\beta^Tk_B^2T\frac{\partial lnZ^T_1}{\partial \left<E_1\right>}
\]
\begin{equation}
\label{entropy-energy-derivative}
\frac{\partial S_1^T}{\partial \left<E_1\right>}=k_B\beta^T
\end{equation}
where we used equations (\ref{temperature-definition}) and (\ref{internal-energy}). This is not a surprising result, since, as was shown in \cite{Plas96,Fra02}, the Legendre transform structure is valid in this context.\\
Now we choose $Y=X_1$. Then:
\[
\frac{\partial S_1^T}{\partial X_1}=\beta^Tk_B^2T\frac{\partial lnZ^T_1}{\partial X_1}.
\]
By equation (\ref{generalized-force}), we conclude that:
\begin{equation}
\frac{\partial S_1^T}{\partial X_1}=-k_B\beta^T\left<\frac{\partial E_1}{\partial X_1}\right>.
\end{equation}
Now, we substitute these last two results in equation (\ref{entropy-differential}):
\[
dS_1^T=k_B\beta^Td\left<E_1\right>-k_B\beta^T\left<\frac{\partial E_1}{\partial X_1}\right>dX_1,
\]
by the definition of the macroscopic work:
\[
dS_1^T=k_B\beta^T(d\left<E_1\right>-\left<dW\right>),
\]
by the first law of thermodynamics:
\begin{equation}
\label{Tsallis-second-law-02}
dS_1^T=k_B\beta^T\left<dQ\right>,
\end{equation}
and, by (\ref{entropy-beta-relation}), we finally arrive at:
\begin{equation}
\label{Tsallis-second-law-01}
TdS_1^T=\left[1-(q_2-1)\beta^*\frac{S_1^T}{k_B}\right]\left<dQ\right>
\end{equation}
At first sight, we see that Tsallis' entropy obeys a modified second law of thermodynamics. This conclusion was also reached by macroscopic considerations \cite{Abe01}. Hence, we can say that the second law is indeed affected by the non-extensive character of the entropy. But we believe that the laws of thermodynamics are above the form of the entropy, they must be valid no matter the context we are working in.\\
We know that the temperature is a integrating factor for the BG entropy. Obviously, we do not have $T$ as the integrating factor for $S_1^T$, from (\ref{Tsallis-second-law-02}) we have the factor $k_B\beta^T$ as the proper integrating factor for the Tsallis' entropy. Hence, we see that no modified second law is needed. We only define a new temperature scale for $S_1^T$, the one set by $k_B\beta^T$.\\
If we retain the temperature as the integrating factor for the second law as the right form of the last, what can we infer from the above relation? From (\ref{Tsallis-second-law-01}) we have, after a simple algebra:
\[
-k_BTd\left[\frac{1}{q_2-1}ln\left[1-(q_2-1)\frac{S_1^T}{k_BT}\right]\right]=\left<dQ\right>
\]
The quantity in brackets is the well known R\`enyi entropy, $S_1^R$. Hence, we conclude that:
\begin{equation}
\label{Renyi-second-law-01}
d(-k_BS_1^R)=\frac{\left<dQ\right>}{T}
\end{equation}
We see that the Renyi entropy satisfies the second law of thermodynamics with the inverse temperature as the integrating factor.\\
From the form of $S_1^R$, we can relate it to the additive variant $S_1$ using equation (\ref{entropy}):
\begin{equation}
\label{Renyi-S1-relation}
S_1=k_BS_1^R+k_B\frac{q_1}{q_1-1}+k_Bln\left[1-(q_2-1)\beta^*\left<E_1\right>\right]^{\frac{1}{q_2-1}}
\end{equation}
Recent papers \cite{MPPl00-01,MPPl00-02,Abe01-01,Abe01-02} also deal with the problem of thermodynamics relations in Tsallis' thermostatistics but using a different approach, namely the $q$-normalized expectation values. The authors of these references obtained equations similar to ours (\ref{internal-energy}) and (\ref{entropy-energy-derivative}). These formulas are different from the ones found here in the fact that they depend on $q$-logarithms, $ln_qZ_1^T$, generalized mean values, $\left<E_1\right>_q$, and the Lagrange multiplier, $\beta^T$. They make a direct connection of such results to the physical temperature, which seems to be unjustified. However, this $q$-normalized formalism is not wrong if we interpret their results as follows: all the relations are correct (which are indeed), and the temperature scale of this $q$-normalized thermodynamics is set by $k_B\beta^T$, as proposed here in equation (\ref{entropy-energy-derivative}) for Tsallis' entropy. Therefore, $S_1^T$ obeys thermodynamics relations of its own, with different mean values, the generalized ones, and temperature scale.\\
Some of them \cite{Len00,Tor02} obtained such relations with natural logarithms of the partition function, but still making use of $q$-averages, the assumption of R\`enyi entropy or macroscopic considerations. The difference from these papers from ours is that we start from a solid physical background: the phase space geometry, where the distribution function is defined.\\

\section{Conclusions}
\noindent
We obtained the microscopic equations for the thermodynamical quantities from the Tsallis' formalism with ordinary mean values. The power law partition function is used and no $q$-logarithm is necessary to obtain such relations. The quantity $lnZ^T_1$ is a natural one.\\
We also argued that no $q$-expectation value is needed to obtain such relations. The issue rests on the fact that the Lagrange multiplier $\beta^T$ is no longer the inverse temperature.\\
The thermodynamical relations here obtained are functions of the inverse temperature $\frac{1}{k_BT}$, which is just the multiplier's value in BG statistics. Hence, Tsallis' formalism can be regarded as an exact canonical ensemble approach and all thermodynamical laws are not changed in this context.\\
The generalized multiplier $\beta^T$ depends on the partition function $Z_1^T$ and sets a new temperature scale for Tsallis' entropy, $S_1^T$. The entropy found here $S_1$ is related to the R\`enyi form $S_1^R$ and, in fact, these two have the physical temperature $T$, as defined by the energy equipartition theorem, as the proper integrating factor in the second law of thermodynamics.\\

\section{Acknowledgments}
\noindent
This work was supported by FUNCAP and CNPq (Brazilian agencies).\\
We we would like to thank prof. M. P. Almeida for many valuable discussions on this matter. We would also like to thank prof. J. E. Moreira for a critical reading of this manuscript.\\

\end{document}